\documentclass[11pt]{article}
\usepackage{times}
\usepackage{sidecap}
\usepackage{geometry}
\usepackage{floatrow}
\usepackage{float}
\usepackage{hyperref}
\usepackage{graphics,graphicx}
\usepackage{subcaption}
\usepackage{epsf,wrapfig}
\usepackage{cleveref}

\crefformat{section}{\S#2#1#3} 
\crefformat{subsection}{\S#2#1#3}
\crefformat{subsubsection}{\S#2#1#3}

\usepackage{natbib,bm}

\usepackage{paralist}
\let\olditem\item
\renewenvironment{thebibliography}[1]{%
  \section*{\refname}
  \let\par\relax
  \renewcommand{\item}[1][]{\olditem[\textbullet]}%
  \inparaenum}{\endinparaenum}

\newcounter{aff}

\DeclareGraphicsExtensions{.jpg,.pdf,.png,.eps}

\usepackage{caption}
\usepackage{upgreek}
\usepackage{xcolor}

\usepackage{tcolorbox}

\bibpunct{(}{)}{;}{a}{}{,}

\usepackage{soul}

\definecolor{darkcyan}{rgb}{0.0, 0.0, 0.7}

\definecolor{boxcolor}{RGB}{193, 229, 255}
\definecolor{figboxcolor}{RGB}{193, 229, 255}
\definecolor{boxcolorgray}{RGB}{227, 227, 227}
\definecolor{darkcyan}{rgb}{0.0, 0.0, 0.7}

\sethlcolor{boxcolorgray}

\newcommand{\msol}{\,M$_{\odot}$}

\def\be{\begin{equation}}
\def\ee{\end{equation}}
\def\bea{\begin{eqnarray*}}
\def\eea{\end{eqnarray*}}

\def\jref@jnl#1{{\rm#1}} 
\def\aj{\jref@jnl{AJ}}			
\def\araa{\jref@jnl{ARA\&A}}		
\def\apj{\jref@jnl{ApJ}}			
\def\apjl{\jref@jnl{ApJ}}		
\def\apjs{\jref@jnl{ApJS}}		
\def\ao{\jref@jnl{Appl.~Opt.}}		
\def\apss{\jref@jnl{Ap\&SS}}		
\def\aap{\jref@jnl{A\&A}}		
\def\aapr{\jref@jnl{A\&A~Rev.}}		
\def\aaps{\jref@jnl{A\&AS}}		
\def\azh{\jref@jnl{AZh}}			
\def\baas{\jref@jnl{BAAS}}		
\def\jrasc{\jref@jnl{JRASC}}		
\def\memras{\jref@jnl{MmRAS}}		
\def\mnras{\jref@jnl{MNRAS}}		
\def\pra{\jref@jnl{Phys.~Rev.~A}}	
\def\prb{\jref@jnl{Phys.~Rev.~B}}	
\def\prc{\jref@jnl{Phys.~Rev.~C}}	
\def\prd{\jref@jnl{Phys.~Rev.~D}}	
\def\pre{\jref@jnl{Phys.~Rev.~E}}	
\def\prl{\jref@jnl{Phys.~Rev.~Lett.}}	
\def\pasp{\jref@jnl{PASP}}		
\def\pasj{\jref@jnl{PASJ}}		
\def\qjras{\jref@jnl{QJRAS}}		
\def\skytel{\jref@jnl{S\&T}}		
\def\solphys{\jref@jnl{Sol.~Phys.}}	
\def\sovast{\jref@jnl{Soviet~Ast.}}	
\def\ssr{\jref@jnl{Space~Sci.~Rev.}}	
\def\zap{\jref@jnl{ZAp}}			
\def\nat{\jref@jnl{Nature}}		
\def\natas{\jref@jnl{Nat.~Astro.}}		
\def\iaucirc{\jref@jnl{IAU~Circ.}}
\def\aplett{\jref@jnl{Astrophys.~Lett.}}
\def\apspr{\jref@jnl{Astrophys.~Space~Phys.~Res.}}
\def\bain{\jref@jnl{Bull.~Astron.~Inst.~Netherlands}}
\def\fcp{\jref@jnl{Fund.~Cosmic~Phys.}}
\def\gca{\jref@jnl{Geochim.~Cosmochim.~Acta}}
\def\grl{\jref@jnl{Geophys.~Res.~Lett.}}
\def\jcp{\jref@jnl{J.~Chem.~Phys.}}	
\def\jgr{\jref@jnl{J.~Geophys.~Res.}}	
\def\jqsrt{\jref@jnl{J.~Quant.~Spec.~Radiat.~Transf.}}
\def\memsai{\jref@jnl{Mem.~Soc.~Astron.~Italiana}}
\def\nphysa{\jref@jnl{Nucl.~Phys.~A}}
\def\physrep{\jref@jnl{Phys.~Rep.}}
\def\physscr{\jref@jnl{Phys.~Scr}}
\def\planss{\jref@jnl{Planet.~Space~Sci.}}
\def\procspie{\jref@jnl{Proc.~SPIE}}

\def\farcs{\hbox{$.\!\!^{\prime\prime}$}}

\def\ion#1#2{#1$\;${\small\rm\@Roman{#2}}\relax}

\newcount\lecurrentfam
\def\LaTeX{\lecurrentfam=\the\fam \leavevmode L\raise.42ex
\hbox{$\fam\lecurrentfam\scriptstyle\kern-.3em A$}\kern-.15em\TeX}
\def\sizrpt{
(\fontname\the\font): em=\the\fontdimen6\font, ex=\the\fontdimen5\font
\typeout{
(\fontname\the\font): em=\the\fontdimen6\font, ex=\the\fontdimen5\font
}}
\definecolor{dark}{rgb}{ 0.34, 0.09, 0.49}

\usepackage{fancyhdr,url}
\usepackage{transparent}
\usepackage{tablefootnote}
\usepackage{mathpazo}

\begin{document}

\noindent
\begin{raggedright}
{\Huge\bfseries Tracing small-scale structure formation through cosmic time with strong gravitational lensing\\[0.5em]
\Large A case for a next-generation sub-millimetre/millimetre interferometer
}\end{raggedright}\\ \smallskip

\noindent

{\raggedright
Hannah R.\ Stacey$^1$, John P. McKean$^{2,3,4}$, Mamta Pandey-Pommier$^5$, 
Carla Cornil-Ba\"{i}otto$^{6}$, Simona Vegetti$^{7}$, 
Aristeidis Amvrosiadis$^{8}$, Fran\c{c}oise Combes$^{9}$, Giulia Despali$^{10,11,12}$, Andreas L.\ Faisst$^{13}$, Giovanni Ferrami$^{7}$, Davide Massari$^{11}$, Dieu D.\ Nguyen$^{14}$, James W.\ Nightingale$^{15}$, Antti Rantala$^{16,17}$, Cristiana Spingola$^{18}$
}

\vspace{1em}
{
\footnotesize{
\noindent $^1$ European Southern Observatory (ESO), Karl-Schwarzschild-Str. 2, 85748 Garching, Germany \\
$^2$ Kapteyn Astronomical Institute, University of Groningen, Postbus 800, NL-9700 AV Groningen, The Netherlands \\
$^3$ South African Radio Astronomy Observatory (SARAO), P.O. Box 443, Krugersdorp 1740, South Africa \\
$^4$ Department of Physics, University of Pretoria, Lynnwood Road, Hatfield, Pretoria 0083, South Africa \\
$^5$ Pole Scientific, Catholic University of Lyon, 10 place des Archives 69002, Lyon, France \\
$^6$ Instituto de F\'{i}sica y Astronom\'{i}a, Universidad de Valpara\'{i}so, Gran Breta\~{n}a 1111, Valpara\'{i}so 2360102, Chile \\
$^7$ Max-Planck-Institut f\"ur Astrophysik, Karl-Schwarzschild-Str. 1, 85748 Garching, Germany \\
$^8$ Department of Physics, University of Oxford, Denys Wilkinson Building, Keble Road, Oxford, OX1 3RH, UK \\
$^9$ Observatoire de Paris, LUX, Coll\`{e}ge de France, CNRS, PSL University, Sorbonne University, 75014, Paris, France \\
$^{10}$ Dipartimento di Fisica e Astronomia {\it Augusto Righi}, Universit\`a di Bologna, via Gobetti 93/2, I-40129 Bologna, Italy \\
$^{11}$ INAF -- Osservatorio di Astrofisica e Scienza dello Spazio di Bologna, via Gobetti 93/3, I-40129 Bologna, Italy \\
$^{12}$ INFN -- Sezione di Bologna, Viale Berti Pichat 6/2, I-40127 Bologna, Italy \\
$^{13}$ IPAC, California Institute of Technology, 1200 E. California Boulevard Pasadena, CA 91125, USA \\
$^{14}$ Department of Astronomy, University of Michigan, 1085 South University Avenue, Ann Arbor, MI 48109, USA \\
$^{15}$ Newcastle University, Herschel Building, Newcastle-upon-Tyne, NE1 7RU, UK \\
$^{16}$ Institute of Astronomy, University of Cambridge, Madingley Road, Cambridge CB3 0HA, UK \\
$^{17}$ Kavli Institute for Cosmology (KICC), University of Cambridge, Madingley Road, Cambridge, CB3 0HA, UK \\
$^{18}$ INAF -- Istituto di Radioastronomia, via Gobetti 101, I-40129 Bologna, Italy
}
\vspace{1em}

}
\begin{tcolorbox}[boxsep=3pt,left=0pt,right=0pt,top=1pt,bottom=1pt,colframe=boxcolor,colback=boxcolor]
\justify
\textbf{Abstract:} 
The nature of dark matter and the growth of supermassive black holes (SMBHs) remain major open questions in modern astrophysics. Both leave a key signature in an overlapping, largely unexplored mass range: $<10^7$\msol\ dark haloes and wandering black holes. These may be undetectable directly, but detectable through gravitational perturbation of strongly lensed images, and partly or entirely distinguishable from one another through lensing data alone. We demonstrate through realistic simulations that a 255-element sub-mm/mm interferometer could probe this critical mass regime, detecting gravitational perturbations in strongly lensed dusty star-forming galaxies (DSFGs) from objects down to at least $10^5$\msol. Combined with lens-galaxy characterisation from the European Extremely Large Telescope (ELT), which strengthens the physical interpretation of detected perturbers, and the unprecedented lens samples expected from {\it Euclid} and the Square Kilometre Array (SKA), this facility could deliver a census of low-mass dark matter haloes and wandering massive black holes across cosmic time.
\end{tcolorbox}

\newpage{}

\vspace{-0.7cm}
\section{Scientific context and open questions}\label{section:science}

\noindent In the standard cosmological paradigm, galaxies form in haloes of non-baryonic dark matter and grow hierarchically through mergers. Two open questions in astrophysics follow directly from this picture: the nature of dark matter and how it forms structures on small scales; and how the seeds of the first massive black holes formed and grew into the SMBHs found at the centre of every massive galaxy today. These questions have profound implications for galaxy evolution \citep[e.g.][]{Somerville:2015,Bullock:2017,Volonteri:2021}. Answering them requires access to a largely unexplored mass range, $<10^7$\msol, where objects at cosmological distances are dark or too faint for existing facilities to characterise in large numbers. As surveys and telescopes in the 2020s and 2030s close in on the bright, high-mass end of galaxy assembly, the discovery-space opportunity in the 2040s lies in the low-mass regime.

\subsection{Low-mass haloes and the nature of dark matter}
\label{section:subhaloes}

Cold dark matter (CDM) is a major component of the $\Lambda$CDM cosmological model, which is currently our best description of the formation and distribution of structures in the Universe. In this model, dark matter consists of low-energy particles that interact only gravitationally. However, this model is not unchallenged, especially at (sub-)galactic scales (\citealt{Bullock:2017} and \citealt{Nadler:2026} for reviews). Moreover, the particle candidates proposed for CDM have not yet been detected. Prominent alternative classes of models include warm dark matter (WDM), ultra-light dark matter (ULDM) and self-interacting dark matter (SIDM), which ascribe different properties to the dark matter particle. While not conclusive, these models also enjoy similar success in observational cosmology (e.g. \citealt{Hui:2017,Adhikari:2025}). 

\begin{figure*}[b!]
    \includegraphics[width=0.84\textwidth]{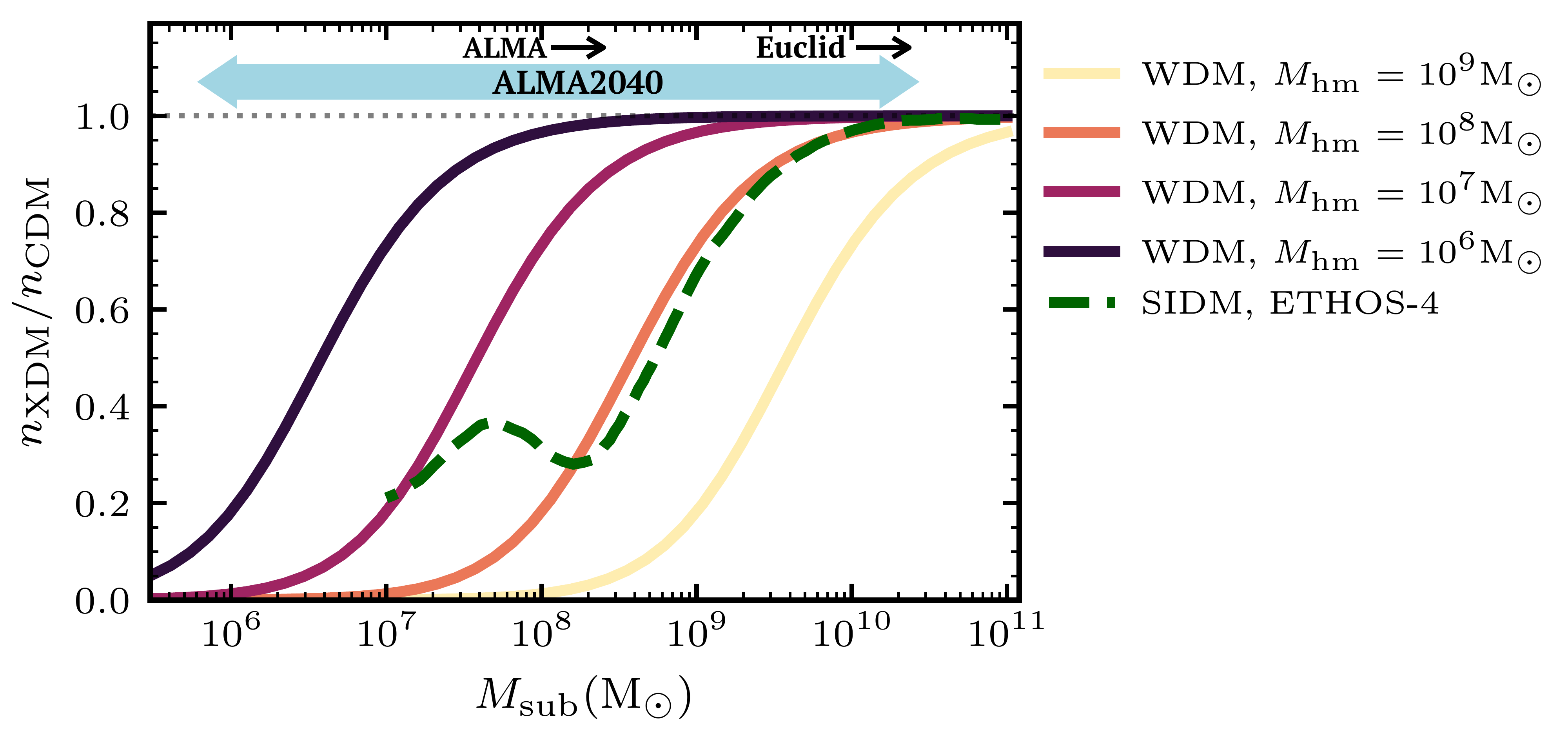}\vspace{-6pt}
    \caption{Ratio of the number of low-mass haloes in alternative dark matter models to that in CDM. The curves show WDM models (\citealt{Lovell:2014}; for various half-mode masses) and ETHOS-4 SIDM simulations \citep{Vogelsberger:2016,Lovell:2020}. A next-generation sub-mm/mm interferometer, such as ALMA2040, could make decisive tests of dark matter by probing low-mass haloes. }
    \label{fig:massfunc}
\end{figure*}

A key difference between these dark matter models is how dark matter is distributed on sub-galactic scales. For example, CDM predicts a large number of low-mass haloes with a well-defined mass function and mass density profile (down to Jupiter-mass scale; \citealt{Wang:2020}). On the other hand, alternative models affect either one of these quantities or both. For example, WDM suppresses both the number density and profile concentration in a way that is directly related to the dark matter particle mass (Fig~\ref{fig:massfunc}; \citealt{Lovell:2014,Ludlow:2016}). SIDM generates more diverse density profiles than CDM \citep{Vogelsberger:2012,Despali:2019}. Therefore, it is possible to discriminate between models of dark matter by finding and characterising low-mass haloes. In order to break the degeneracy between the effects induced by changes to the physics of dark matter and that of baryons, it is essential to probe dark matter haloes below $10^7$\msol\ which are expected to be devoid of stars and gas. Without a detectable electromagnetic signature, these objects can only be characterised via their gravitational influence.

\subsection{Wandering black holes and primordial seeds}
\label{section:BHs}

All massive galaxies are believed to contain SMBHs. Analytic and semi-analytic models of SMBH growth suggest that small seed black holes form early, then grow through gas accretion and hierarchical mergers as galaxies assemble. However, the nature of the initial seeds remains unsettled. Potential seed candidates include Population~III remnants, heavy seeds from direct collapse, and medium-weight seeds from dense star clusters, which predict relic populations of $10^{3}$ to $10^{6}$\msol\ black holes \citep{Mezcua:2017,Greene:2020,Volonteri:2021}. 

Cosmological simulations further predict that some of these black holes never settle to galaxy centres: minor mergers and satellite stripping deposit them in wide orbits, and they wander in the halo \citep{Ricarte:2021b}. Some may also be recoiling out from the nucleus in dynamical interactions \citep{Bekenstein:1973}. However, models make diverse predictions about their numbers that heavily depend on numerical prescriptions for seeding, accretion, feedback, and dynamical friction \citep{Greene:2020}. Estimates at the high end predict massive galaxies ($M_{200}\sim10^{13}$\msol) may host 100s of wanderers $>10^5$\msol \citep{Ricarte:2021b,DiMatteo:2023}. Therefore, a census of $<10^6$\msol\ wandering black holes across redshifts could discriminate between different seeding channels (especially between star cluster and direct collapse seeds) and probe sinking efficiency, while $>10^8$\msol\ recoiling black holes probe late-stage pre-merger dynamics. However, the accretion duty cycles and Eddington ratios of most wandering black holes are far too low for electromagnetic detection, leaving rare tidal disruption events (e.g. \citealt{Stein:2026}) the only route to constrain the quiescent wandering population via an electromagnetic signature, while pulsar timing and gravitational wave observatories probe only the merging binary population \citep{Amaro-Seoane:2023}.

\subsection{Gravitational detection of low-mass objects}
\label{section:gravitational}

\noindent Strong gravitational lensing by galaxies and galaxy clusters provides a uniquely sensitive probe of mass, whether dark or baryonic. It has therefore become a powerful diagnostic of the nature of dark matter, particularly because it can reveal low-mass haloes (both sub-haloes and line-of-sight structures) via perturbations to the surface brightness of lensed images (see \citealt{Vegetti:2024} for a review). This sensitivity to mass perturbations makes strong lensing, in principle, capable of detecting sub-haloes (\cref{section:subhaloes}) and wandering black holes (\cref{section:BHs}) directly via their gravitational lensing signature.

\begin{figure*}[t!]
    \centering\vspace{-4pt}
    \includegraphics[width=0.67\textwidth]{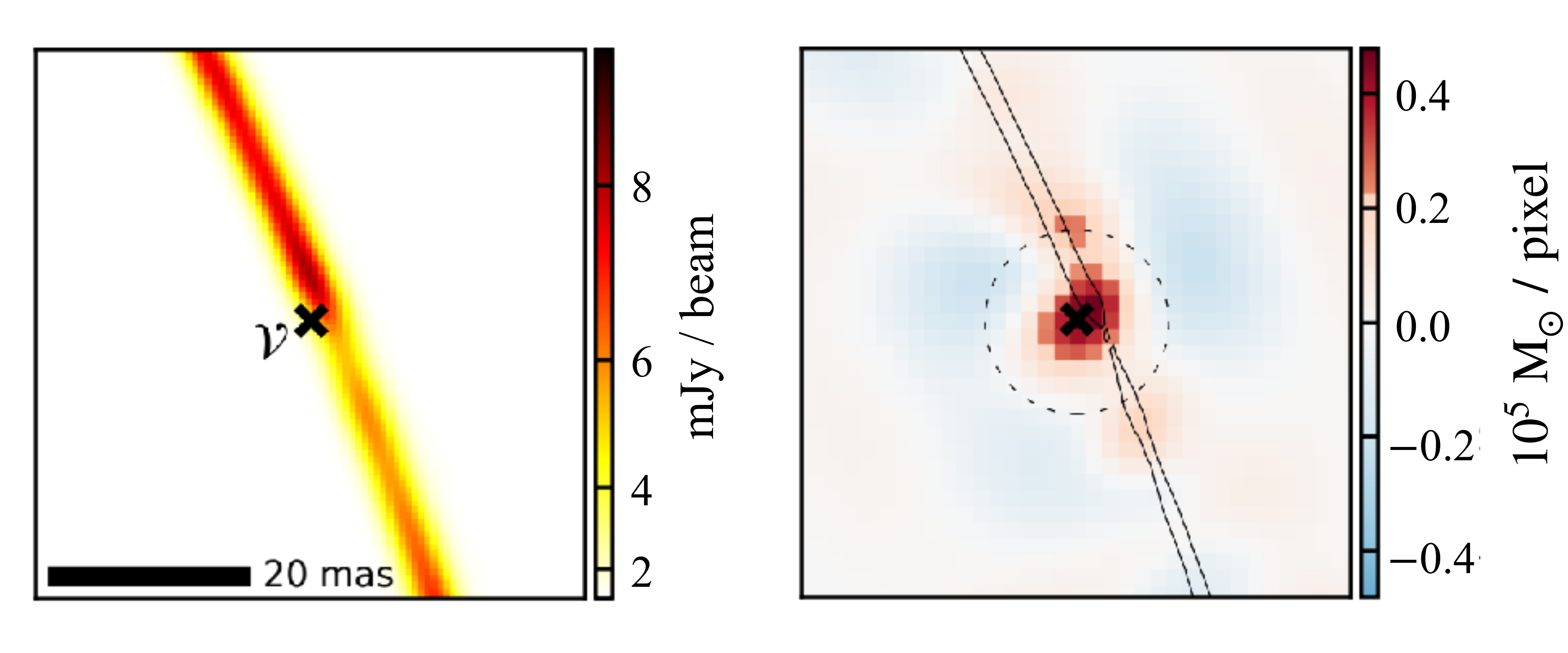}\vspace{-4pt}
    \caption{Detection of a $(1.13\pm0.04) \times 10^6$\,\msol\ object in a gravitationally lensed arc with $\approx5$~milliarcsecond angular resolution using cm-VLBI observations \citep{McKean:2025,Powell:2025}. Left: discontinuity in the lensed arc caused by the perturber ($\nu$). Right: pixelated surface mass density corrections revealing the compact perturber. A combination of milliarcsecond angular resolution and high surface brightness sensitivity is required to probe objects in this largely unexplored mass regime. Figures adapted from \citet{Powell:2025}.}
    \label{fig:powell2025}
\end{figure*}

Current high-resolution, near-infrared imaging (e.g. JWST/NIRCam) of galaxy- and cluster-scale strong lenses is sensitive to low-mass haloes in the mass range $10^8$ to $10^{10}$\msol\ (for median CDM mass concentrations; e.g. \citealt{Despali:2022,Nightingale:2022}). Such data have resulted in three statistically robust detections \citep{Vegetti:2010,Vegetti:2012,Powell:2025} and constraints on dark matter cosmologies \citep[e.g.][]{Vegetti:2014b,Vegetti:2018,Vegetti:2026,Ritondale:2019}. Several studies have shown that these gravitationally detected low-mass haloes can be simultaneously constrained in redshift, mass and density profile \citep{Minor:2021,Minor:2025,Enzi:2025,Tajalli:2025,Despali:2025}, with their results ranging from marginally to fully inconsistent with CDM. Recently, \cite{Tajalli:2026} showed mathematically that arcs with sufficient signal-to-noise ratio and thickness, observed with the required angular resolution, can constrain not only the mass and density profile but potentially also the curvature and higher-order derivatives of the deflection profile.

The current Atacama Large (sub-)Millimetre Array (ALMA) can potentially push to lower masses ($>10^7$\msol; \citealt{Stacey:2025b}) through observations of the large population of starburst galaxies identified in mm/sub-mm surveys \citep{Negrello:2010,Vieira:2013}. Current studies are limited by the paucity of bright targets and ALMA's angular resolution, given the current sensitivity and baseline configuration. However, a crucial test of dark matter is detecting haloes in the $10^6$ to $10^7$\msol\ range because they differ dramatically in abundance and density profiles between dark matter models (Fig.~\ref{fig:massfunc}). These haloes have a lensing signature (deflections and Einstein rings) with image separations of the order of a few milliarcseconds.

\cite{Powell:2025} presented the current state-of-the-art in this field, using milliarcsecond interferometric imaging at cm-wavelengths \citep{McKean:2025} to detect a perturber with a well-defined mass of just  $10^{6.05\pm0.02}$\msol\ within a projected radius of 80~pc (Fig.~\ref{fig:powell2025}). This study demonstrates that objects in the crucial mass regime of $\sim10^6$\msol\ are detectable, provided the data have sufficient sensitivity and angular resolution, and that the required complex analysis techniques are already mature \citep{Powell:2021}. Furthermore, an analysis of the density profile of the same perturber by \cite{Vegetti:2026} shows that the high-quality lensing data alone can discriminate between the various physical scenarios presented in \cref{section:science}: a point mass (describing a wandering black hole), a King profile (describing a globular cluster), and a single NFW profile (a dark-matter-dominated halo). These models were each strongly disfavoured relative to a two-component model of an unresolved $10^{5.53\pm0.03}$\msol\ point mass within $\leq10$\,pc, centred on an extended disc structure of size 140\,pc. This combined structure appears to be in tension with CDM and is only tentatively accommodated by SIDM \citep{Vegetti:2026}.

While the results for the three detected perturbers provide a powerful test of dark matter, both the novelty of the detected density profiles and the uniqueness of the results in the crucial $\leq10^7$\msol\ mass range underscore how a statistically significant sample is essential to rule out any single model conclusively. Furthermore, these results come with the caveat that they required exceptionally well-constrained detections, which highlights both the need for high-quality data and the benefit of complementary infrared data from ELT to constrain any baryonic component (see \cref{section:ELT}). The key issue is that the number of such detections remains low across optical, infrared, mm and radio wavelengths, and a much larger sample of lens systems needs to be observed with sufficient depth at milliarcsecond resolution.

\begin{figure}[t!]
    \centering\vspace{-2pt}
    \includegraphics[width=0.83\textwidth]{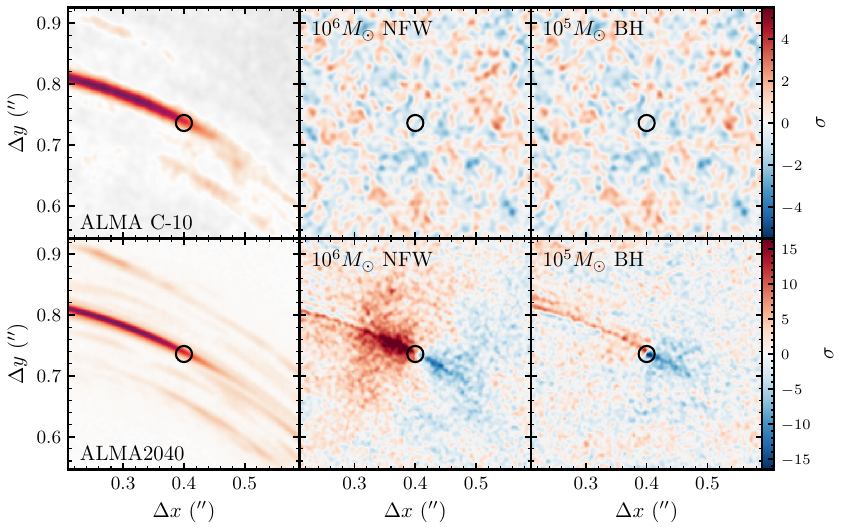}\vspace{-3pt}
    \caption{Simulated observations for ALMA C-10 and ALMA2040 of a lens with a $10^6$\msol\ dark matter halo and a $10^5$\msol\ black hole placed on an arc at the lens redshift, $z=0.5$. Left: zoomed-in deconvolved image with natural weighting; the circle shows the perturber location. Middle and right: differences between dirty images of mock data with and without the perturber in units of the rms noise ($\sigma$). \cref{section:simulations} gives details on how we generated the mock data.}
    \label{fig:subhalo}
\end{figure}

\section{Technical requirements for a next-generation sub-mm interferometer}
\label{section:technical}

\subsection{Requirements for detecting low-mass perturbers}
 \label{section:requirements}

\noindent Multiple factors play into the ability to detect and characterise low-mass perturbers via lensing. Angular resolution and surface brightness sensitivity are both important. Contamination from lens light at optical and near-infrared wavelengths \citep{Nightingale:2022}, and the structure and magnification of the lensed sources \citep{Koopmans:2005,Despali:2022,Tajalli:2026} are also important factors. Finally, useful statistics are essential; therefore, a significant sky volume depends on the number of lens systems, the size of lensed arcs, and the distance to the lensed source. In all these aspects, a next-generation sub-mm/mm interferometer that could succeed ALMA (hereafter, ALMA2040) offers unique advantages and complements other forthcoming facilities.

\begin{figure}[!t]
    \centering\vspace{-2pt}
    \includegraphics[width=0.78\textwidth]{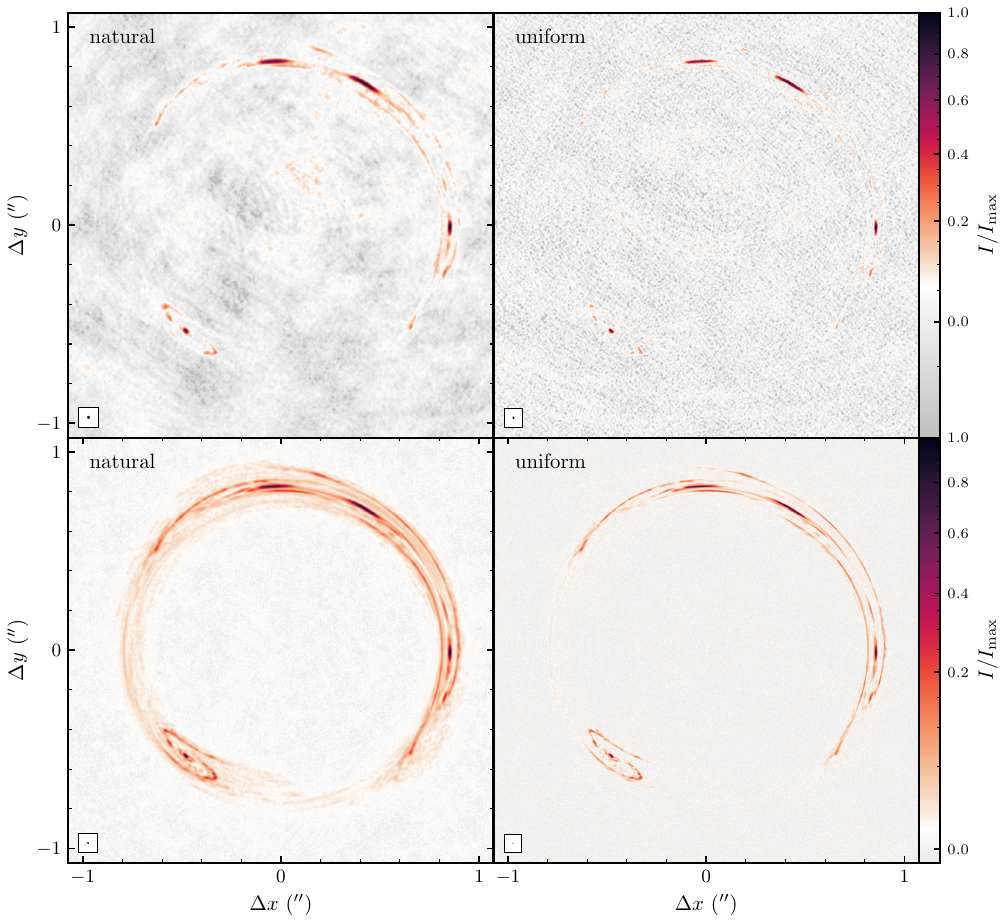}\vspace{-4pt}
    \caption{Deconvolved images of simulated data of a gravitational lens using the current ALMA C--10 configuration (top row) and using a potential ALMA2040 configuration of 255 antennas (bottom row). Both simulations are one hour of on-source integration. The colour maps use a square-root stretch. The beam size is shown in the box in the lower-left corner.}
    \label{fig:simulations}
\end{figure}

ALMA observations on 10s of pc-scales of the sub-mm dust emission from strongly lensed dusty star-forming galaxies (DSFGs) reveal compact structures suitable for dark matter lensing studies \citep{Stacey:2024}, and, given their larger extent, they probe a larger sky volume per lens than cm-VLBI (e.g.~Fig.~\ref{fig:powell2025}). In addition, even as sensitivity improves in the era of SKA-VLBI and the ngVLA, there may not be more targets with extended emission that can be detected with cm-VLBI. While the number of lensed objects will increase, the non-thermal radio emission may remain compact around the central engine of less powerful active galaxies and not produce extended arcs needed for gravitational detection. 

However, ALMA is at the limit of what is achievable without a major sensitivity improvement. More than an order-of-magnitude step in sensitivity would significantly increase the sky volume in which low-mass perturbers could be detected, which can be obtained through a larger collecting area and instantaneous bandwidth. Currently, lensed DSFGs are identified from flux-limited samples with relatively large thresholds \citep{Negrello:2010,Vieira:2013}. Cross-matching of lenses from wide-area surveys (e.g. {\it Euclid} or JWST) with low-resolution sub-mm/mm surveys can also reveal new sub-mm-bright lenses, which can be followed up efficiently with ALMA2040 sensitivity, at the cost of lower purity. This would increase the parent sample of lensed objects available for low-mass substructure studies by an order of magnitude and allow better target selection. Also, baselines 2 to 3 times as long as current ALMA could obtain 5 milliarcsecond angular resolution at the sensitive frequency range around current ALMA Bands 6 and 7.

Fig.~\ref{fig:subhalo} demonstrates the influence of a $10^6$\msol\ NFW dark matter halo and a $10^5$\msol\ black hole placed on an extended gravitational arc in a simulated ALMA2040 observation (details of the simulations are given in \cref{section:simulations}). Both low-mass objects create $\geq15\upsigma$ significance distortions in surface brightness relative to the perturber-free simulated data. Such high-resolution observations will also enable constraints on the objects' density profiles, which, together with their abundance, are crucial to discriminate between dark matter models (e.g. \citealt{Tajalli:2025}) and, in principle, between a point-mass black hole or star cluster and an extended dark matter halo of comparable total mass. 

\begin{wrapfigure}[21]{b}{0.33\textwidth}\vspace{-7pt}
    \includegraphics[width=\textwidth]{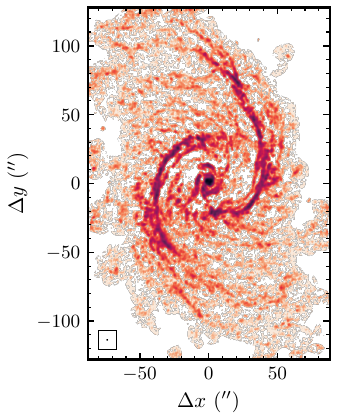}\vspace{-6pt}
    \caption{Image of integrated CO line emission from nearby galaxy NGC\,1566 used for the lensed source \citep{Leroy:2021}. Shown with log-stretch. Beam is shown in the lower-left corner.}
    \label{fig:real}
\end{wrapfigure}

Three further advantages set ALMA2040 apart from current (JWST) and future (ELT) optical and near-infrared alternatives. Emission from the foreground lens (particularly stellar, but also dust and AGN) is negligible or undetectable in the sub-mm regime, as most lenses are low-redshift early-type galaxies, meaning that contamination from lens light complexity does not meaningfully limit lensing studies. In rare cases, a foreground AGN can be excluded using the spectral index. In contrast, this is likely to be problematic for future optical/infrared instruments such as ELT, as the push for greater sensitivity reveals more low-surface-brightness foreground emission that blends with lensed emission. Despite both DSFGs and strong lenses being intrinsically rare on the sky, lensed DSFGs are much easier to identify in the sub-mm in large-area surveys compared to the optical/infrared \citep{Negrello:2010,Vieira:2013}, including lensed sources that will be out of reach of the ELT; the dozens of already identified lenses, alone, may be enough to exclude CDM or its alternatives. Finally, bright emission lines such as [CII] can be detected out to very high redshifts: the 3D information provides more stringent constraints than collapsed 2D emission \citep{Hezaveh:2013b}, which would enable independent constraints on perturber properties.

\subsection{Simulated data}
\label{section:simulations}

We created realistic simulated observations of a gravitational lens with a 255-antenna interferometer to demonstrate improved angular resolution and dynamic range relative to ALMA. The source is based on ALMA observations of CO\,(2--1) from the nearby Seyfert galaxy NGC\,1566 \citep{Leroy:2021} obtained from the PHANGS collaboration public database\footnote{\url{https://www.canfar.net/storage/vault/list/phangs/RELEASES/PHANGS-ALMA}} (see Fig.~\ref{fig:real}). The image of NGC\,1566 uses combined ALMA, ACA and Total Power data to cover a large range of angular scales. We assume the integrated CO line image of NGC~1566 as dust continuum for our simulations, assuming that the spatial distribution of redshifted sub-mm dust emission and CO is similar. While the true properties of NGC\,1566 may not represent a real high-redshift sub-mm galaxy, we use it to include compact and extended structures at angular resolutions beyond current observations. The spiral structure resembles real lensed galaxies, such as HERS\,1 \citep{deRoo:2025}. We do not simulate any emission from the lens, as foreground emission is not expected at 350~GHz.

We scale the source model to a projected size of $\approx 2$\,kpc at $z_{\rm src}=4$ to match the typical extent of sub-mm galaxies (e.g. \citealt{Stacey:2025a,deRoo:2025}). We used PyAutoLens \citep{Nightingale:2021} to lens the source by a singular isothermal ellipsoid (SIE) density profile at $z_{\rm lens}=0.5$ with an Einstein radius of $0\farcs8$ plus external shear (similar to the lens model for SPT\,0538--50 determined by \citealt{Stacey:2024}). We scaled the model to the total flux density of SPT\,0538--50 in real ALMA observations with a similar central frequency and 30~milliarcsecond angular resolution, as reported by \citet{Stacey:2025a}. We also created models with a $10^6$\msol\ NFW (assuming a mass-concentration relation from \citealt{Ludlow:2016}) and $10^5$\msol\ point mass placed on an arc at the same redshift as the lens.

We generated simulated observations using the CASA simulator toolkit \citep{CASATeam:2022}. Table~\ref{table:simulations} summarises the simulation parameters. We assume an instantaneous bandwidth of 128\,GHz\footnote{While even wider bandwidths may be considered, the atmospheric absorption towards shorter wavelengths means there will be little gain in sensitivity, while longer wavelengths would reduce the angular resolution.} and an antenna configuration with a maximum baseline length of 28\,km. All antennas are 12\,m in diameter. The sensitivity assumes no gain from improvements in antenna, receiver, or signal-chain efficiency. While the simulated antenna configuration and bandwidth imply a factor of 20 improvement in surface-brightness sensitivity, we find a factor of 40 improvement (not accounting for the higher angular resolution), which we attribute to the better $uv$ coverage of the ALMA2040 simulation. 

We show images of the mock data created using CLEAN for visualisation in Fig.~\ref{fig:simulations}. However, recent developments in lens modelling techniques already enable full forward-modelling of visibility data (e.g. \citealt{Powell:2021,Stacey:2024,Stacey:2025a,Stacey:2025b,McKean:2025,Amvrosiadis:2025}).

\section{Synergies with current and future facilities}

\subsection{\textbf{\textit{Euclid} and \textit{Rubin}/LSST}}
\label{section:Euclid}

The ongoing ESA {\it Euclid} mission is on the way to detecting an expected $10^5$ strong lens systems by the end of its six-year survey \citep{EuclidCollaboration:2026}. {\it Euclid}'s own strong-lensing analysis may report some low-mass halo detections, but at masses comparable to the existing near-infrared work described in \cref{section:gravitational} ($>10^{8.8}$\msol; \citealt{ORiordan:2025}). Meanwhile, the Vera C. Rubin Observatory recently began its 10-year Legacy Survey of Space and Time (LSST), which will probe the dynamic Universe by continuously monitoring 18000 deg$^2$ of sky \citep{Ivezic:2019} and discover a similar number of galaxy-scale strong lenses to {\it Euclid} \citep{Shajib:2025}. Thousands of these will be lensed quasars, which are typically hosted in dusty star-forming galaxies \citep{Stacey:2018} and could provide additional sub-mm-bright targets for ALMA2040.

While {\it Euclid} and LSST can provide information on dark matter via the number and density profiles of more massive haloes, their role is primarily to supply the large, well-characterised parent lens population from which sub-mm-bright, structurally suitable sources can be selected (\cref{section:requirements}) to build a statistically significant sample. In addition, they will provide information on the baryonic properties of the lenses, which includes the stellar light profile of the main halo and the position of luminous satellites, and will also provide independent constraints on the lens mass model from a different source surface brightness distribution. 

\subsection{SKA}
\label{section:SKA}

The currently under-construction SKA is a next-generation radio interferometer that will operate between 50 and 350 MHz (SKA-Low; Australia) and between 350 MHz and 15.4 GHz (SKA-Mid; South Africa). At the highest frequencies, the SKA will provide sensitive (few $\upmu$Jy~beam$^{-1}$), wide field-of-view (0.5~deg$^2$) and high angular resolution (few 100\,mas) imaging well-suited to lens finding \citep{Rezaei:2022}. The SKA is expected to discover around $10^5$ new gravitationally lensed radio sources \citep{McKean:2015,McCarty:2025,Pandey-Pommier02.2026.SKA}. 

The SKA, operating as a VLBI facility, could help identify low-mass structure directly, as in the case of \cite{Powell:2025}. However, its main synergy will likely be supplying a large number of viable targets over a matched sky area with ALMA2040. This works because radio emission from radio-faint lensed sources is dominated by star-formation processes \citep{Bonzini:2013}, and the well-known correlation between radio synchrotron and far-infrared dust emission \citep{Bell:2003} then allows these sources to be reliably cross-matched against ALMA2040's sub-mm/mm targets.

\subsection{ELT}
\label{section:ELT}

The forthcoming ELT will provide an angular resolution comparable to ALMA2040 \citep{Padovani:2023}. While this may yield perturber detections in the optical and near-infrared, it will depend strongly on lens light contamination and PSF quality. In this case, ELT/MICADO imaging could provide an independent, achromatic check of perturber detections in combination with ALMA2040. Furthermore, the ELT will identify foreground lenses and lens shapes that can help refine the mass model and improve model constraints, which is essential for group- and cluster-scale lenses. 

However, the main strength of ELT lies in identifying luminous counterparts. In this sense, ALMA2040 and ELT complement each other: follow-up strengthens perturber classification and breaks degeneracies in perturber characteristics (e.g., redshift vs density profile), while ALMA2040 lensing data provide a mass diagnostic for galaxies where dynamical mass measurements from ELT spectroscopic data are infeasible. 

\subsection{LISA}
\label{section:LISA}

The Laser Interferometer Space Antenna (LISA), planned for launch around 2035, will detect gravitational waves from merging massive black hole binaries across cosmic history \citep{Amaro-Seoane:2023}. This probes a different population than that of ALMA2040: LISA will be sensitive to the final inspiral stage of black hole mergers in the $10^4$ to $10^7$\msol\ mass range, while ALMA2040 will be sensitive to seeds that have not (or may never) merged. Those with dynamical friction timescales that exceed Hubble time ($>10^7$\msol) will never be detectable by gravitational-wave observatories or pulsar timing arrays. These probes are therefore complementary. A census of wandering black holes from ALMA2040 would constrain both the merging and non-merging seed population, providing key insight into their formation mechanisms.\par
\medskip\noindent
Together, the synergies described in \cref{section:Euclid} to \cref{section:LISA} will allow lensing analyses with ALMA2040 to draw from unprecedented lens samples ({\it Euclid} and SKA), anchor a robust mass model and classify perturber detections (ELT), and situate the census of wandering black holes within the broader context of black hole assembly (LISA). Together, they provide an observational picture of dark matter structures and black hole growth that can only be delivered with a next-generation sub-mm/mm interferometer.

\begin{table}
    \caption{Summary of the simulated continuum data.}
    \centering \vspace{-7pt}
    \begin{tabular}{l|cc}
         & ALMA C--10 & ALMA2040 (ext.) \\ \hline
       Phase centre & \multicolumn{2}{c}{J2000 19h40m00s $-$40d00m00s} \\
       Total integration & \multicolumn{2}{c}{1h} \\
       Central frequency  & \multicolumn{2}{c}{350\,GHz} \\
       Bandwidth  & 7.5~GHz & 128~GHz\footnote{We scale up to the larger bandwidth assuming a constant source surface brightness and noise} \\
       $uv$ maximum & 16~km & 28~km \\
       Beam (natural) & $0\farcs015\times0\farcs014$ & $0\farcs009\times0\farcs009$ \\
       Beam (uniform) & $0\farcs012\times0\farcs010$  & $0\farcs005\times0\farcs005$ \\
       rms noise (natural) & 28\,$\upmu$Jy\,beam$^{-1}$ & 0.7\,$\upmu$Jy\,beam$^{-1}$ \\
    \end{tabular}
    \label{table:simulations}
\end{table}

\subsection*{Acknowledgements}
We would like to thank Manuel Aravena and Chris Fassnacht for their helpful comments and careful review of the manuscript. This work was developed as part of the broader ALMA2040 community effort. We acknowledge the Lorentz Center and NOVA for supporting the community workshops and coordination activities that helped initiate and shape the ALMA2040 White Paper process. JPM is supported in part by the National Research Foundation of South Africa (Grant number: SARCI170317224762 / 128943). DDN acknowledges support from the ELT Postdoctoral Research Fellowship at the Department of Astronomy, University of Michigan. JWN is supported by an STFC/UKRI Ernest Rutherford Fellowship, Project Reference: ST/X003086/1. CS acknowledges the support of the Italian Ministry of University and Research (grant FIS2023-01611, CUP C53C25000300001).\medskip

\raggedright
\bibliography{references2} 
\bibliographystyle{apj_mod}
\end{document}